# The Barolo Palace:
# Medieval Astronomy in the Streets of Buenos Aires

## Alejandro Gangui

**Abstract.** Cultural heritage relating to the sky in the form of sundials, old observatories and the like, are commonly found in many cities in the Old World, but rarely in the New. This paper examines astronomical heritage embodied in the Barolo Palace in Buenos Aires. While references to Dante Alighieri and his poetry are scattered in streets, buildings and monuments around the Western world, in the city of Buenos Aires, the only street carrying Dante's name is less than three blocks long and, appropriately, is a continuation of Virgilio street. A couple of Italian immigrants—a wealthy businessman, Luis Barolo, and an imaginative architect, Mario Palanti—foresaw this situation nearly a century ago, and did not save any efforts or money with the aim of getting Dante and his cosmology an appropriate monumental recognition, in reinforced concrete. The Barolo Palace is a unique combination of both astronomy and the worldview displayed in the Divine Comedy, Dante's poetic masterpiece. It is known that the Palace's design was inspired by the great poet, but the details are not recorded; this paper relies on Dante's text to consider whether it may add to our understanding of the building. Although the links of the Palace's main architectural structure with the three realms of the Comedy have been studied in the past, its unique astronomical flavor has not been sufficiently emphasized. The word of God, as interpreted by the Fathers of the Church in Sacred Scripture, Aristotle's physics and Ptolemy's astronomy, all beautifully converge in Dante's verses, and the Barolo Palace reflects this.

### Medieval Cosmology and the World of Dante

Dante Alighieri, who became famous for his *Commedia*, a monumental poem written roughly between 1307 and his death in 1321, and which the critics from sixteenth century onwards dubbed *Divina*, is among the most noteworthy poets of Western culture. In this and other works, Dante describes the cosmic image of the world, summing up the current trends of Neoplatonic and Islamic traditions.[1] His universe consisted of a fixed

---

1 Miguel Asín Palacios, *La escatología musulmana en la Divina Comedia*, (4th ed.) (Madrid: Hiperión, 1984), pp. 133–34; James Dauphiné, *Le cosmos de Dante* (Paris: Les Belles Lettres 1984), pp. 13–14.

Earth, in the center of the cosmos, surrounded by nine 'heavens' or spheres. Traveling from the inner sphere towards the outer, as Dante himself experiences during his dream-vision trip in Paradise, we find the heavens of the Moon, Mercury, Venus, the Sun, Mars, Jupiter, Saturn, the heaven of the fixed stars and the metaphysical sphere of the prime (or unmoved) mover (the Aristotelian *primum mobile*).

Dante's ideas of the cosmos are well—and elegantly—expressed in the *Commedia*, but more so in his unfinished encyclopedic philosophical treatise *The Banquet* (*Il Convivio*). Other hints to his vast scientific learning are also scattered in his other works, for example in *The New Life* (*La Vita Nuova*), his first collection of lyrics expressing his love for Beatrice.

There are a number of useful accounts of Dante's universe by Dreyer, Dauphiné and Cornish, on which much of the following description of Dante's cosmology is based.[2] For readers familiar with Spanish language, some handy astronomical notes can be found in *Poética Astronómica: El Cosmos de Dante Alighieri*.[3] Medieval cosmology, including Dante's cosmology prevailed for many centuries and holds a continued fascination, For example, C. S. Lewis wrote that 'I have made no serious effort to hide the fact that the old [medieval] Model delights me as I believe it delighted our ancestors. Few constructions of the imagination seem to me to have combined splendor, sobriety, and coherence in the same degree'.[4]

**Dante's Architecture of the Afterlife**

---

The universe of Dante is described by medieval Christian theology and Ptolemaic astronomy. It consists of a simplified Aristotelian model, where the sphere of the Earth is standing still at the geometric center of the universe. The northern hemisphere is the only habitable land where humanity can live, and it has Mount Zion at its center—at its navel—near Jerusalem. Ninety degrees to the East, one finds the river Ganges; ninety degrees to the West, the Ebro in the Iberian Peninsula. The southern hemisphere has no dry land emerging from the immense ocean covering it and is therefore forbidden to man. However, a huge mountain lies just at its center, exactly opposite to Mount Zion. Both rivers and both mountains inscribe a cross within the sphere of the Earth.[5] Below Zion, with the form of an inverted cone, is Hell, the place where sinners are punished. It is divided into nine decreasing concentric circles which culminate at the center of the Earth, and therefore also of the universe. From there, there is a crack, or narrow tunnel, opened up by the waters of the river Lethe, the river of forgetfulness in Hades, which connects the bottom of Hell with the base of the mountain of Purgatory, in the opposite hemisphere. This second realm of the afterlife, according to Dante's verses, is thus located in the midst of the unexplored ocean of the uninhabited hemisphere directly opposite to Jerusalem. And, should the souls reach the top of the mountain, they would be cleansed of all sin and made perfect, and therefore wait their turn to ascend to Heaven.

During the Middle Ages, there was a strong desire to develop a realistic model that described, as closely as possible, the actual structure of the world. In the worldview devised by Dante, the mountain of Purgatory—as we may imagine it—was so high that its third terrace laid above the Earth's atmosphere and its summit nearly reached the sphere of fire. In 'La Divina Comedia', Jorge Luis Borges writes: 'we tend to believe that he [Dante] imagined the other world exactly as he presents it [...] that Dante imagined that, once he was dead, he would find the inverted mountain of Hell, the terraces of Purgatory or the concentric heavens of Paradise [...] Obviously, this is absurd'; Borges then recalls the commentary of Dante's son on the *Divine Comedy*: 'he said his father intended to show the sinner's life through the image of Hell, penitent's life through the image of Purgatory and the life of righteous souls via the

---

image of Paradise. He did not read it in a literal way'.[6] The Garden of Eden was located on top of this mountain and, surrounding the Earth, there were several solid, but transparent, spheres which carried the astronomical heavens. These nine spheres were nested one inside the other and, represented the heavens of the Moon, Mercury, Venus, the Sun, Mars, Jupiter, and Saturn; encasing these all, there came the immense sphere of the fixed stars and, still going outwards, the metaphysical invisible sphere of the *primum mobile*.[7]

In his vision of the *Commedia*, Dante, the pilgrim, travels through this world, first in company of the Roman poet Virgil, who leads Dante around the sloping sides of the punishment realm. Traversing circles—nine in total—of gradually decreasing sizes, their journey depicts the worst sinners located deepest on this conic-shaped Hell, until they reach the apex of the cone, at the exact center of the Earth. After having passed the bottom of the abyss—where the king of the fallen angels, Lucifer, dwells—they continue their trip, but now re-ascending through an underground passage, to the other side of the spherical Earth. Having survived, as Borges puts it, 'la suciedad, la tristeza y el horror del Infierno', or 'the dirty, the sadness and the horror of Hell', Dante and his guide escape from the underworld, to the mountain of Purgatory.[8] Together they traverse two lower slopes—called 'ante-Purgatory'—at the base of the mountain, and then climb the seven terraces of the mount, which correspond to the seven deadly sins that, one at a time, ought to be purged by the souls. Adding all these together, one finds, again, the ubiquitous number nine, this time for the distinct regions of this second realm of the afterlife. Having passed the terraces and reached the earthly paradise at the top of Purgatory, Dante is finally permitted to start his journey through the heavens. But he needs a new guide. Virgil, as a pagan, is not allowed to enter Paradise, and must remain forever in Limbo, the first circle of Hell. Here Dante meets Beatrice and she will be his new companion through the celestial realm.

**An Eclectic Project: Palanti and Barolo Joined by Dante**

---

6 Jorge L. Borges, 'La Divina Comedia' (1980), in: *Siete noches*. Published in *Obras completas*, tomo III (1975–85) (Barcelona: Emecé, 1996), p. 207.
7 Cornish, 'Dante's moral cosmology', p. 203.

8 Borges, 'La Divina Comedia', p. 210; my translation.

The Italian architect Mario Palanti (1885–1979) was in charge of the works of the Italian Pavilion for the Argentine Centenary Exhibition of 1910. There are no official biographies of Mario Palanti and no written records of his intentions regarding the Barolo Palace, apart from the very details of the Palace construction, and a few mentions by Hilger.[9] After returning to his home country and serving in the First World War, in 1919 he traveled back to Argentina. This time, Palanti reaches the southern hemisphere with one single idea in mind: to materialize a votive offering, in the guise of a monumental Temple, honouring Dante in the 600[th] anniversary of his death. In the past, such temples were made in order to gain favour with supernatural forces, which may be the very same forces that, Palanti imagined, propelled Dante along his trip (Figs. 1–3). The temple had to be located in the ascensional axis of the souls dwelling in Purgatory and Paradise, in the antipodes of Mount Zion, where the fortress conquered by King David was placed, in Jerusalem. In Buenos Aires, Palanti met Luis Barolo and the project took shape.

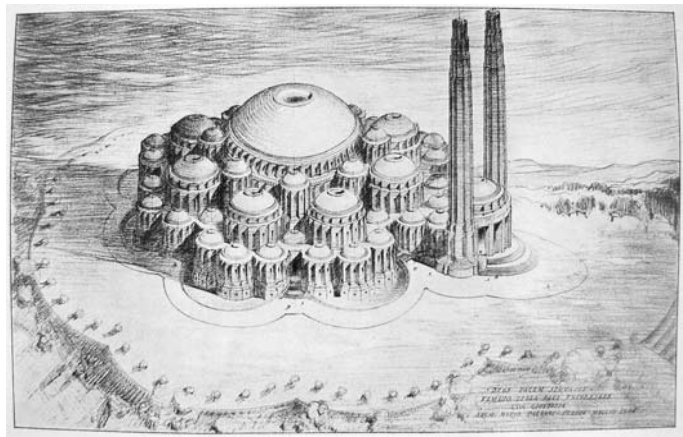

**Figure 1**: One of the many projects conceived by the Italian architect Mario Palanti. 'Temple of the universal peace. Initial perspective study of the great work planned out', in Mario Palanti: *Architettura per tutti* (Milano: Emilio Bestetti, 1947).

9 Carlos Hilger, 'Monumento al genio latino', *Summa+* 3 (October-November 1993), p. 37; Carlos Hilger, 'Capriccio italiano', pp. 35–45, in: C. Braun and J. Cacciatore, eds., *Arquitectos europeos y Buenos Aires: 1860–1940* (Buenos Aires: Fundación TIAU, 1996).

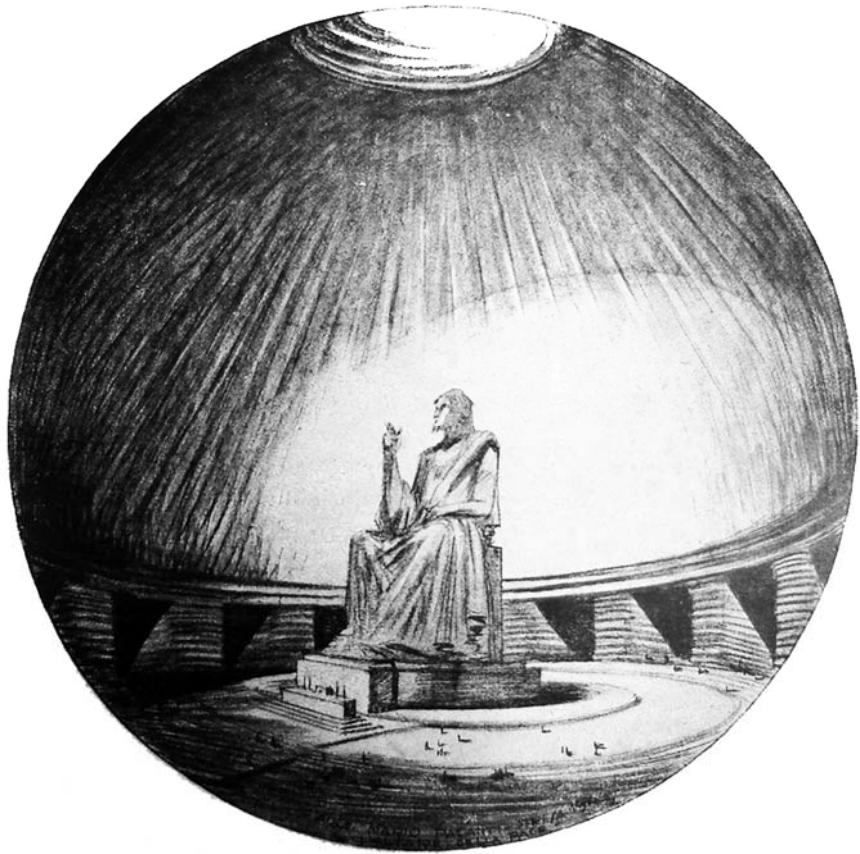

**Figure 2**: 'Temple of the universal peace. Christ Redeemer blessing mankind. Monumental statue inside the room of the highest dome. At the base the peace altar', in Palanti, *Architettura per tutti*.

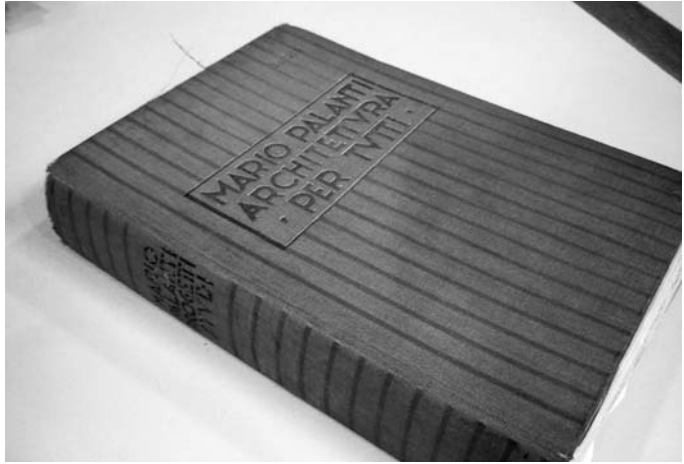

**Figure 3**: Mario Palanti's book *Architettura per tutti* (Milano: Emilio Bestetti, 1947). Limited bilingual edition (English-Italian). This copy, number 1919, is in the National Library of Argentina. It was catalogued in year 1949.

Luis Barolo (1869–1922) was an Italian businessman and the wealthy sponsor that Palanti needed in order to implement his dream. Like Palanti, Barolo was a lover of his native country, Italy, and of Dante Alighieri. Also like Palanti, there are no official biographies of Barolo and no written records of his intentions regarding his Palace, the only information being recorded by Hilger. We do know that the only wool-spinning mill in Argentina belonged to him, and that his cashmeres were first-rate. Moreover, like many immigrants, Barolo was worried about international military conflicts and the necessity of preservation of European culture in the face of the disasters of the war. Having met Palanti at the peak of his business success, Barolo agreed to finance Palanti's project aiming to immortalize Italian culture and Dante's verses (Fig. 4).

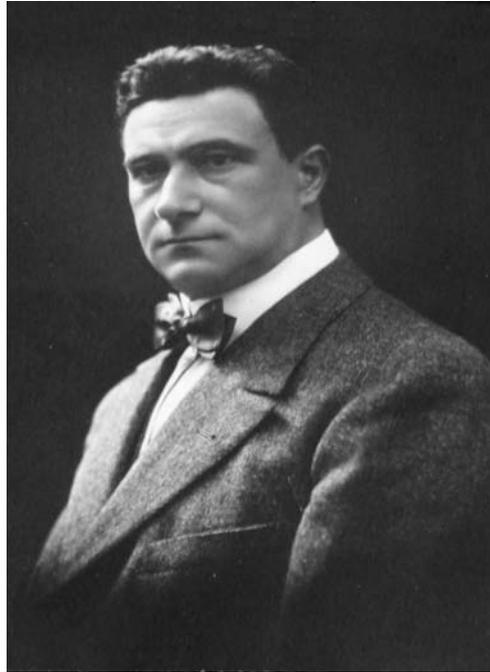

**Figure 4**: The Italian businessman Luis Barolo.

As previously mentioned, Palanti wanted his future monumental building to be located in a place diametrically opposite to Mount Zion (Jerusalem). But that antipodal point was very far away from Buenos Aires. Let us recall that Dante and Virgil, after having escaped from the abyss of Hell through a natural underground passage (the *natural burella*), emerged on the shores of an island located in the southern hemisphere, where they found the mountain of Purgatory. Now, Jerusalem's coordinates are roughly 31°47′N, 35°13′E, so that its antipodes fall approximately at the latitude of Buenos Aires, 34° 36′S, but completely missing the city's longitude, 58° 20′W. Indeed, the mythic mountain, according to topographical representations of the death realms depicted for Dante's journey, ought to be located at longitude 144°47′W, which clearly makes the famous mountain fall in the middle of the Pacific Ocean.

But this fact did not inhibit Palanti. The architect may have recalled Dante's verses on Ulysses' last voyage to the southern hemisphere when, endangering his sailors with the aim to gain knowledge of the unknown, the hero from Ithaca traverses those Pillars 'where Hercules his landmarks set as signals,/ That man no farther onward should adventure'.[10] These are the famous verses where Ulysses gives strength to his companions by recalling the origin (the seed) of the human race, its moral virtues and the virtue of the intellect (the pursuit of knowledge): 'Considerate la vostra semenza:/ fatti non foste a viver come bruti,/ ma per seguir virtute e canoscenza' ('Consider ye the seed from which ye sprang;/ Ye were not made to live like unto brutes,/ But for pursuit of virtue and of knowledge'.[11] And so Palanti imagines a new set of fabulous *Pillars of Hercules*, this time in the estuary of the River Plate. Thus, he conceives the construction of votive monuments on both sides of the River Plate: the Salvo Palace, to be located in Montevideo, and the Barolo Palace on the shores of Buenos Aires.

**The Barolo Palace**
The Barolo Palace was conceived as the materialization—as closely as possible—of the *Divine Comedy*. As such, it is full of allegories and symbolism referring to Dante's work. For example, the building is divided vertically into three parts, exactly as the many pictorial 'representations' depict Dante's Hell, Purgatory and Paradise, the three realms of the afterlife. Palanti designed the building so that it became a scale model of Dante's universe, following the tradition of the Gothic cathedrals, the supreme achievement of stonework in the Middle Ages. And so, as the main door of the Paris cathedral represents an alchemical rite of passage and that of Chartres depicts an astrological reference book also the Barolo Palace was built with a deep—and explicit—symbolisms in mind.[12]

The arched walls of the ground floor passage constructed by the architect put together a series of vaults, with a height equivalent to three standard levels, and which are filled with Latin inscriptions, such as one from Virgil's *Eclogues*, 'trahit sua quemque voluptas' (Each man is led by his own taste); others written also by Virgil are: 'sic vos non vobis nidificatis aves' (Thus, do ye birds build nests for others), and 'sic vos

---

10 Inf. XXVI, 108–09.

11 Inf. XXVI, 118–20.

12 See Hilger, 'Monumento al genio latino', p. 40.

non vobis mellificatis apes' (Thus, do ye bees make honey for others).[13]
But one can also find quotations from Horace, Ovid and from biblical
writings as well. As it was usual of Palanti's eclectic constructions the
palace was conceived as a lay temple for the promotion of liberal arts
(Figs. 5 and 6).[14]

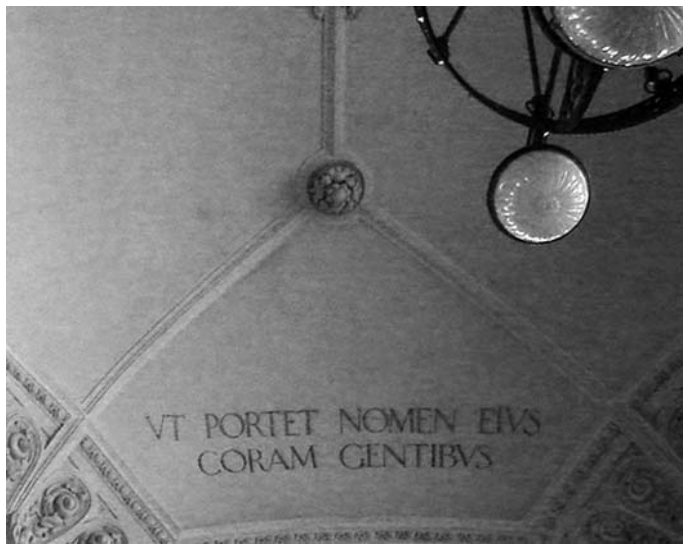

**Figure 5**: Latin inscription in the arched walls of the ground floor passage of
the Barolo Palace. 'Ut portet nomen elus coram gentibus' (Bear my
name before the nations and kings, *New Testament*, *Acts of the
Apostles* 9:15). In this phrase, 'gentibus' comes from *gentile*, and
refers to *all* the people, not just the religious ones, children of
Israel.

---

13 Virgil's *Eclogues*, II, 65.

14 Mario Palanti, *Architettura per tutti* (Milano: Emilio Bestetti, 1947).

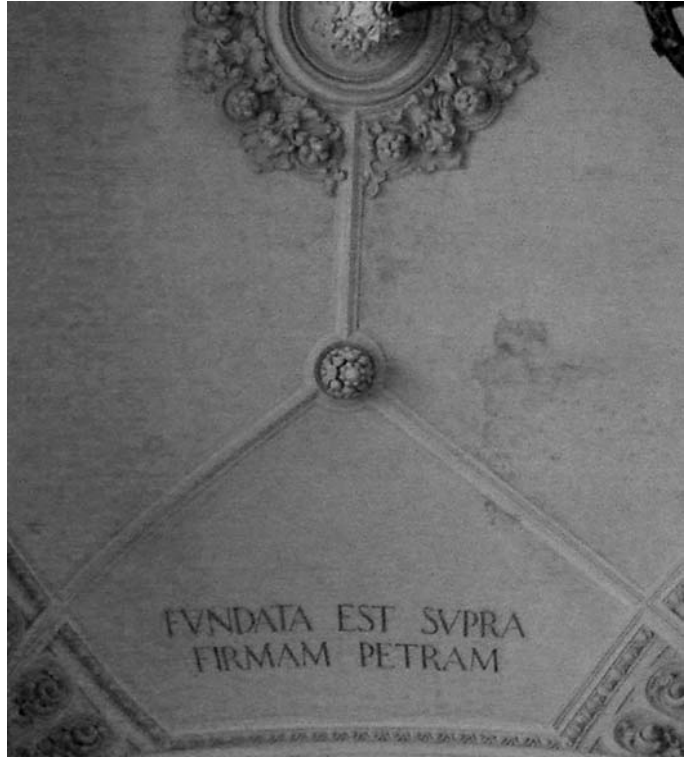

**Figure 6**: Latin inscription in the arched walls of the ground floor passage of the Barolo Palace. 'Fundata est supra firmam petram' (This is the house of the Lord solidly built, it is well founded on solid rock, *New Testament*, *Matthew* 7:25).

The building itself turned out to be monumental, considering the time when it was erected, of course: it comprises 22 floors and 2 basements, reaching a height of approximately 100 meters. This detail was not by chance: the *Divine Comedy* comprises exactly 100 cantos, which for Dante represented the 'perfect number' (*i.e.*, 10) multiplied by itself. The city code of the time, however, limited the maximum height allowed for new buildings along May Avenue (*Avenida de Mayo*) in the city of Buenos Aires, where the Palace is located, and so a special permission had to be granted by the City Mayor, José Luis Cantilo. *The Barolo*, as it is now known, set quite a few 'firsts', being considered the first high building of the city, and indeed the tallest in Latin America in its time. It continued being the highest of Argentina until the construction of the Kavanagh building in 1936, and it was the first entirely made of reinforced concrete, a technique with which Palanti had good experience, but which had no precedents in Argentina by that time.[15]

When one leaves the ground level and its vaults and walks up the stairs, on the third floor one enters the cleansing realm of Purgatory. From the third floor to the 14th, one goes up—one 'terrace' after the other—as Dante did along the mountainside. Then, after the 14th, one can climb up more seven stories through the tower of the Palace, a clear allegory of the seven deadly sins to be purged in each of the terraces of Purgatory. At the very top, the Palace is crowned with a rotating lighthouse, the material representation of the angelic choruses of Dante's Paradise. This contains a voltaic arc lamp of roughly 300,000 *bougies* (approximately 300,000 candelas, equivalent to around 2,500 ordinary incandescent light bulbs). All these numbers combined in the sort of numerological portrait of reality so appreciated by Pythagoreans, and followed also by Dante, gives: $14 + 7 + 1 = 22$. By dividing this result by 7 (seven deadly sins or stories through the tower), yields 22/7 which is the simplest Diophantine approximation of number Pi, the ratio of a circle's perimeter to its diameter.[16] Again, there is no coincidence in this; Palanti apparently twisted his design to make the circle—the most perfect geometric figure—ciphered within the Palace's architecture.

Of course, taking this 'realistic' iconography—and especially Dante's own words—one cannot rightly say that Purgatory lies 'on top of' Hell, for, as we saw above, these two realms are 'located' in opposite hemispheres of the Earth. Dante 'the writer' suggests this when, having arrived to the center of the Earth and descended along the body of giant Lucifer (the 'fell worm who mines the world'), he abruptly gets upside down after passing a certain point. Surprised, Dante 'the pilgrim' asks Virgil to explain to him what had happened, getting as a reply: 'Thou still imaginest/ Thou art beyond the centre, where I grasped/ The hair of the fell worm, who mines the world./ That side thou wast, so long as I

---

15 Palanti, *Architettura per tutti*; Mario Palanti, *Palandomus* (Milano: Lucini, n.d.).

16 Ángel J. Battistessa, 'Introducción', in: *Dante Alighieri, La divina comedia: Infierno* (Buenos Aires: Asociación Dante Alighieri, 2003), p. 35.

descended;/ When round I turned me, thou didst pass the point/ To which things heavy draw from every side'.[17]

Dante and his guide had just gone through that special 'point' onto which all weights converge in the search for their natural place, in the very sense Aristotle conceived it. Having passed to the southern hemisphere they then stopped descending and began climbing to reach the surface opposite to Jerusalem: 'We mounted up, he first and I the second,/ Till I beheld through a round aperture/ Some of the beauteous things that Heaven doth bear;/ Thence we came forth to rebehold the stars'.[18] The pilgrims are now ready to start the second part of their journey in Purgatory.

So strictly speaking, Purgatory is not 'above' Hell (within Dante's architecture of the world, one can even claim that Lucifer has both his head and his legs pointing 'upwards'; Fig. 7). The physical partition of the Barolo Palace, however, allegorically needs Hell to be below all the rest, and this first realm of Dante's trip is represented by the ground floor. So Palanti envisioned Hell as the main entrance of the Palace, as it was the case for Dante when he went astray (allegorically) in the dark wood, at the very entrance of Hell (Inf. I, 1–3). And, like the hollow void of Hell which is divided into nine infernal circles of punishment, also the central passage of the Palace (the pedestrian alley which connects two parallel streets, on the ground floor level) has nine vaults of access: three towards *Avenida de Mayo*, which is the main entrance of the Palace; three towards the former *Victoria* street (now *Hipólito Yrigoyen* street), parallel to *Avenida de Mayo*; one given by the central vault, which stretches out towards the dome; and two other ones that contain the staircases, on both sides of the central vault (Figs. 8 and 9).

---

17 Inf. XXXIV, 106–11.

18 Inf. XXXIV, 136–39.

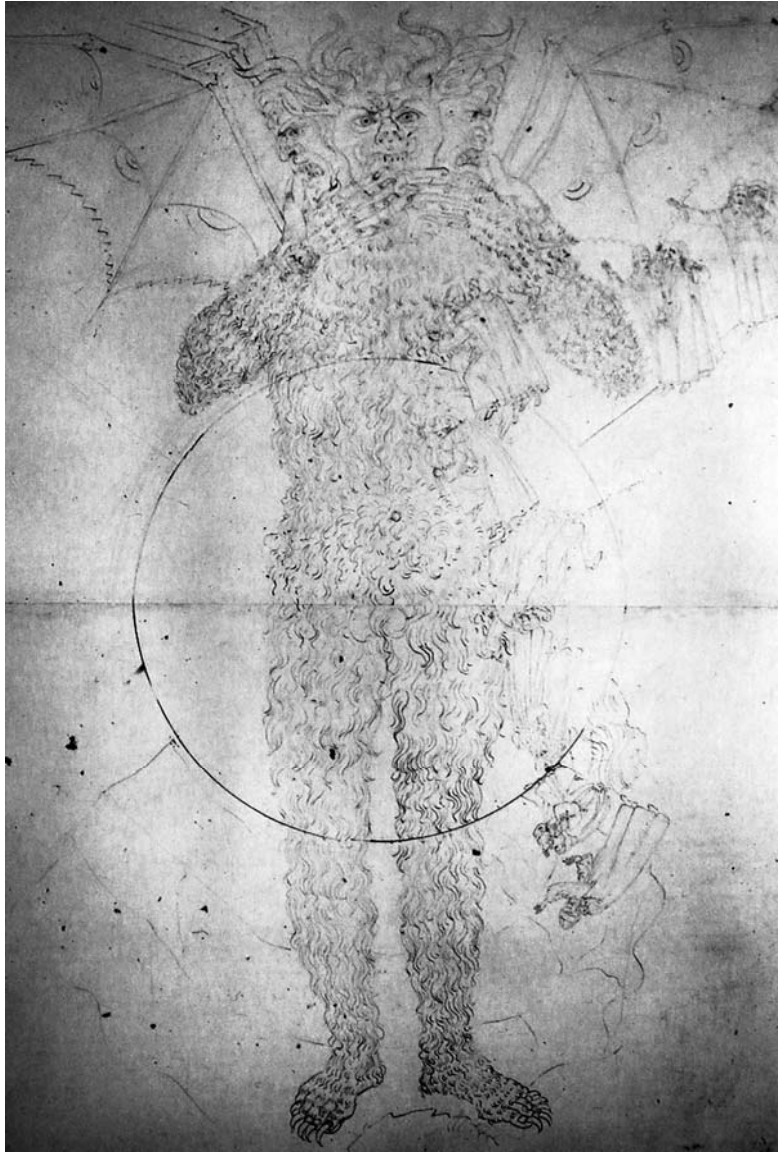

**Figure 7**: Sandro Botticelli's drawing for the *Commedia*, depicting one scene of the last canto of the *Inferno* when Dante and Virgil descended along the giant Lucifer's body, who was stuck in the center of the Earth and, suddenly, got turned upside down (facsimile from the National Library of Argentina).

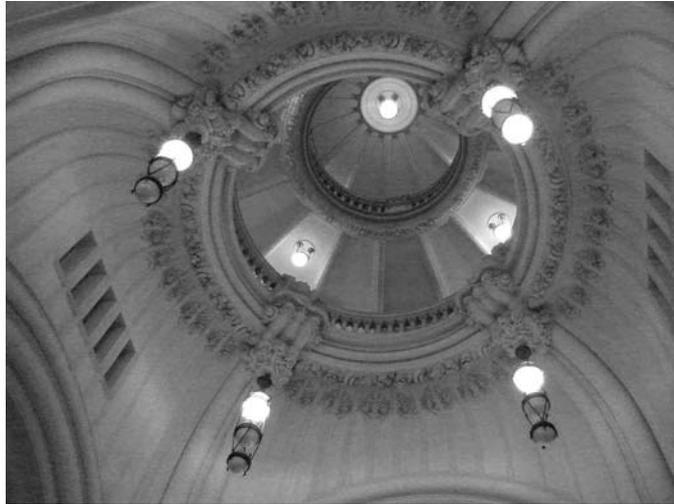

**Figure 8**: Central vault as seen from the ground floor of the Barolo Palace.

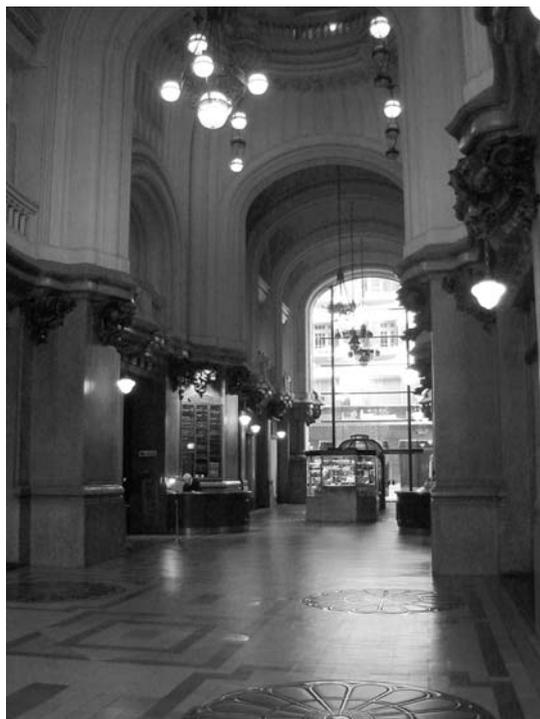

**Figure 9**: Central passage of the Palace. At the end of the pedestrian alley one can see the exit towards *Hipólito Yrigoyen* street.

**The Number Nine**
The number 'nine' is ubiquitous in many of Dante's works, in addition to the *Commedia*. For example, the following text from *La Vita Nuova*:

> Nine times already since my birth the heaven of light had circled back to almost the same point, when there appeared before my eyes the now glorious lady of my mind, who was called Beatrice even by those who did not know what her name was. She had been in this life long enough for the heaven of the fixed stars to be able to move a twelfth of a degree to the East in her time; that is, she appeared to me at about the beginning of her ninth year, and I first saw her near the end of my ninth year.[19]

It has been suggested that Dante was employing the precession of the equinoxes to tell us Beatrice's age when they first met.[20] To follow this line of reasoning, let us remember that the value Dante considered for precession was roughly 36,000 years (and not 25,765 years). So the actual computation that Dante did was: 360 degrees correspond to 36,000 years, from which it follows that 1/12 of a degree ('a twelfth of a degree') corresponds to approximately 8 years. When Dante says: 'she appeared to me at about the beginning of her ninth year', he means that Beatrice was finishing her eighth year of age when they first met.

In the whole text of *The New Life* the number nine appears on ten occasions. For example: 'After so many days had passed that precisely nine years were ending since the appearance, just described, of this most gracious lady'[21]; 'It was precisely the ninth hour of that day, three o'clock in the afternoon, when her sweet greeting came to me'[22]; and 'At

---

once I began to reflect, and I discovered that the hour at which that vision had appeared to me was the fourth hour of the night; that is, it was exactly the first of the last nine hours of the night'.[23] Dante also related the number nine to Beatrice's premature death, which occurred on June 8[th] 1290, when she was 24. Dante described it as follows: 'she departed in that year of our Christian era (that is in the year of Our Lord) in which the perfect number had been completed nine times in that century in which she had been placed in this world: she was a Christian of the Thirteenth Century'.[24] A possible interpretation of this is as follows: 1290 = 10 ('the perfect number') x 9 ('completed nine times') + 1200 ('in that century in which she had been placed in this world: she was a Christian of the Thirteenth century'). In the same chapter Dante describes how 'according to Ptolemy and according to Christian truth, there are nine heavens that move', and also that 'this number [*i.e.*, number nine] was in harmony with her'.[25] He continues 'it will be clear that this number was she herself' and concludes simply that 'she was a nine, or a miracle, whose root, namely that of the miracle, is the miraculous Trinity itself'.[26] There is therefore ample evidence that the number nine, and its root, three, characterize both Dante's cosmic architecture and his allegorical descriptions of it.

**Climbing Barolo's Purgatory: Turning Right or Left?**
Returning to the Barolo Palace, the two staircases on both sides of the central vault, are symmetric. Hence, one of these requires the passerby to ascend, and turn to the right heading for the second part of the building, namely Purgatory, while the other turns to the left. In the *Commedia* there is a preferred way for the pilgrims to descend through the circles of Hell, downwards to the center of the Earth, and then climb upwards along the terraces of Mount Purgatory (Figs. 10–12).

---

23 Vita Nuova III, 8.

24 Vita Nuova XXIX, 1.

25 Vita Nuova XXIX, 2.

26 Vita Nuova XXIX, 3.

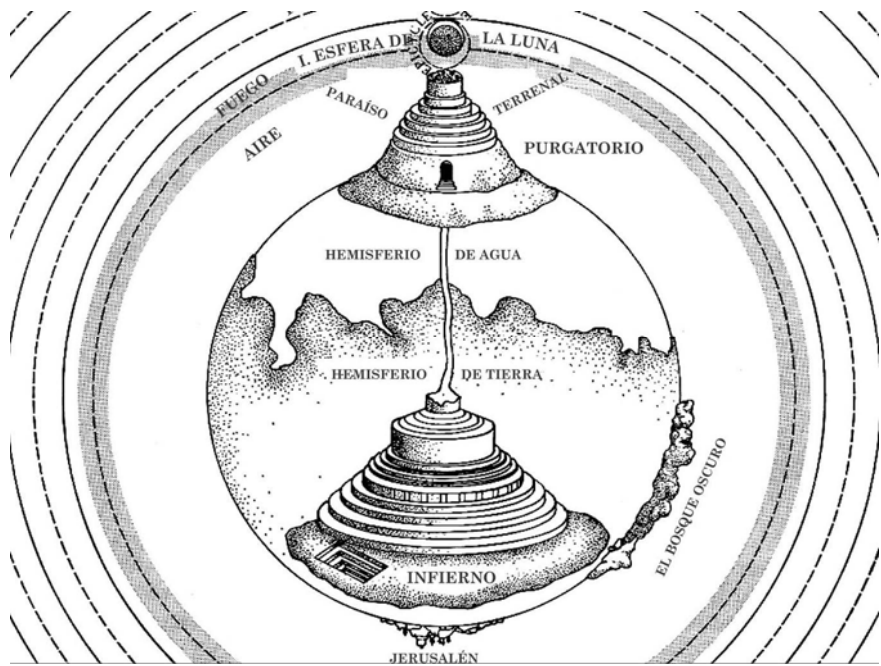

**Figure 10**: Purgatory and Hell according to one 'realistic' topographical representation of the realms traversed during Dante's journey. Pilgrims descend along circles of the concave walls interior to the cavity of Hell, spiraling to the left. The ascent along the convex path of the terraces of Mount Purgatory winds up to the right. (Image adapted from Charles Singer, ed., *Studies in the History and Method of Science*, Vol. I [1917], Fig.4).

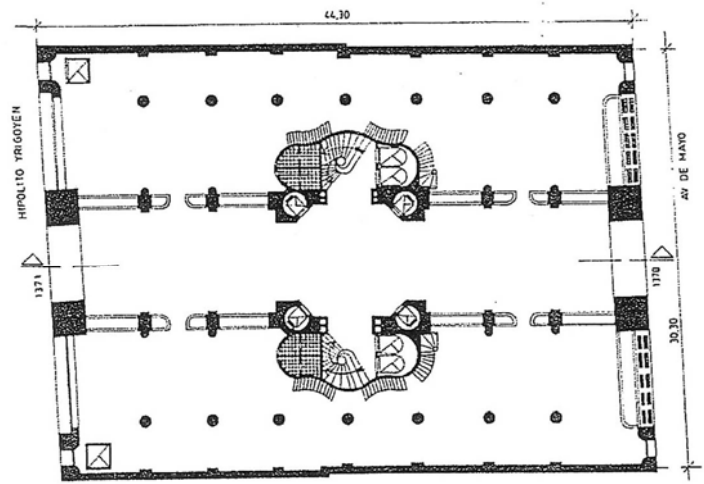

Planta baja

**Figure 11**: Plan of the ground floor of the Barolo Palace, showing the two main staircases on both sides of the central vault (Solsona and Hunter, *La Avenida de Mayo*).

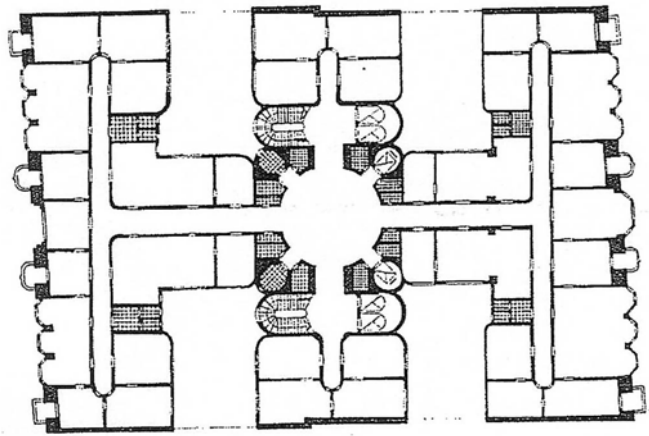

Planta tipo

**Figure 12**: Plan of a typical story of the Barolo Palace. The right hand side border of the plan faces May Avenue (*Avenida de Mayo*); it is the main façade of the building (Solsona and Hunter, *La Avenida de Mayo*).

The usual representations of Dante's cosmos show the location on Earth of two of the three landmarks of the afterlife, Hell and Purgatory. Originating in a violent and sudden cataclysm, both are cone-like and complementary, as the first is a hollow void while the second is a solid mass. The pilgrim's descent proceeds along the circles of the concave walls interior to the cavity of Hell, which spiral down to the left, whereas the ascent passes along the convex path of the terraces of Mount Purgatory winding up to the right. Given the inverse placement of both realms, the trajectory of the pilgrims is always in the same rectilinear direction towards the top of Mount Purgatory and the heavenly spheres, the Empyrean and, finally, towards God.[27]

This orientation during Dante's journey is not arbitrary, but conforms to a scriptural tradition in which right means superior and left, inferior. In *Genesis* (13:9), due to the lack of grass and water to raise his sheep, Abraham went to the right and his nephew Lot to the left. Hopper points out that Christ ordained that the sheep should stay at his right hand side, while the goats should remain on his left.[28] The same tradition occurs in the *Koran* (sura 69, ayat 13–31), which reads 'when the trumpet shall sound one blast [...] On that day ye will be exposed; not a secret of you will be hidden. [...] as for him who is given his record in his right hand [...] he shall be in a life of pleasure, in a lofty garden' But, 'as for him who is given his record in his left hand [...] Take him and fetter him! And burn ye him in the Blazing Fire!' Similar hints are found in the *Odyssey* in which, Dante observed, Ulysses, after traversing the Pillars of Hercules, sailed to the south-west: 'Evermore gaining on the larboard side' that is, sailing ever to the left; after a journey of five months, he saw a mountain that, in his words, 'seemed to me so high/ As I had never any one beheld' and, soon afterwards, his ship sank and the sea above them 'closed again', killing everybody.[29]

It seems consistent with the literary tradition with which Dante would have been familiar, that he chose the path on the left when he was traversing Hell, but changed to the path on his right hand side during his journey along the terraces of Purgatory.

---

27 Cornish, 'Dante's moral cosmology'.

28 Vincent F. Hopper, *Medieval Number Symbolism: Its Sources, Meaning, and Influence on Thought and Expression* (New York: Columbia University Press, 1938).

29 Inf. XXVI, 126.

However, it is clear that not both staircases of the Barolo building—as they were built by Palanti, and can still be seen in the detailed drawing to scale of the construction—can satisfy the requirement of spiraling up and to the left, for the first two floors, and then up and to the right, for the rest of the main building. Tourist guides might then suggest visitors to take the appropriate staircase—first, one of them and, on the third floor, the opposite one—when the elevators (nine in total, as it should be) are not operating properly.

**Syllables and Modules**

The ubiquitous importance of the number nine, and of its root, the number three, in Dante's work has been observed. Each of the three canticas in which the *Divine Comedy* is divided, contains 33 cantos (plus one as an introduction, which makes up 100 cantos in total). Dante knew of the tradition that Christ was aged thirty-three at the time of his death and resurrection.[30]

The literary metric employed by Dante is constructed with verses with an enchained rime, called *terza rima* (aba, bcb, cdc, ...). The verse here is hendecasyllabic, meaning that the last stress in placed on the tenth syllable. It will therefore contain 11 syllables if the verse has *la uscita piana* (that is, if after the tenth syllable it follows another one without stress), as it often happens in Italian poetry.

It appears that Palanti also wanted to incorporate this characteristic feature of the *Commedia* in the Palace, and so he divided each story in the building's main body of the Barolo Palace into 11 modules, which are apparent in the 11 windows and balconies of the Barolo front façade (Fig.13).

---

30 Battistessa 'Introducción', p. 35.

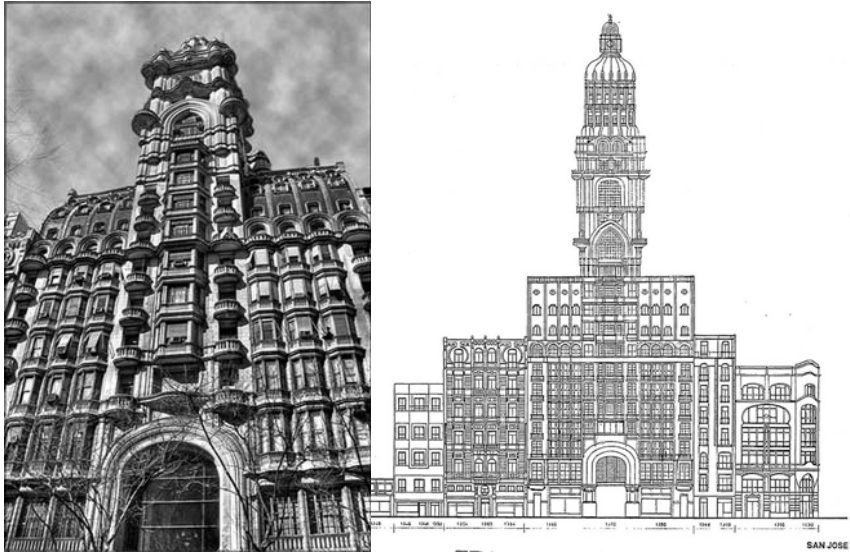

**Figure 13**: Left, Front façade of the Barolo Palace today. The main entrance of the building faces *Avenida de Mayo*. Right, Sketch of the front façade of the Barolo Palace found in Solsona and Hunter, *La Avenida de Mayo*, 1990.

**The Doors of Paradise**

From the Garden of Eden—the earthly paradise on top of Mount Purgatory—located just below the Aristotelian sphere of fire, it is Beatrice who guides Dante along the last leg of his journey, through the nine celestial spheres of Heaven. Virgil, as a pagan, was forbidden access to Paradise and was forever obliged to remain in Limbo, the first circle of Hell.[31] Paradise is the abode of angels and saints, and was structured, according to Judeo-Christian tradition, by the classical nine concentric, spherical and incorruptible, heavens: the eight astronomical Ptolemaic 'spheres' plus the Aristotelian *primum mobile*.[32] Surrounding this all, there was the Empyrean, as the tenth and last theological sphere, infinite and motionless, made just of light and beyond space and time, as its dwelling was in the mind of God.

---

31 Inf. IV, p. 39.

32 Edward Grant, *Planets, Stars and Orbs: The Medieval Cosmos 1200–1687* (Cambridge: Cambridge University Press, 1996).

Dante and Beatrice pass successively from the heaven of the Moon to that of Saturn, at which they are ready to enter the sphere of the stars. It is at this moment that Dante gives a clue to his probable date of birth, writing that he saw the constellation following Taurus, namely Gemini. Then, suddenly, faster than when you put 'thy finger in the fire and drawn it out again', he found himself inside of this new constellation.[33] In the following verses, he sings to 'his' stars: 'O glorious stars, O light impregnated/ With mighty virtue, from which I acknowledge/ All of my genius, whatsoe'er it be,/ With you was born, and hid himself with you,/ He who is father of all mortal life,/ When first I tasted of the Tuscan air'.[34] Dante is telling us with these verses that the Sun, 'father of all mortal life', rose and set in conjunction with the constellation Gemini, when Dante breathed the air of Tuscany for the first time. From his own words, then, one can narrow Dante's birth date to between May 22 and June 21 (in the modern calendar), 1265, approximately.

So, Gemini seems to be Dante's own constellation and birth sign, and it is from it that he afterwards carries on with his trip through the heavens. In a sense, we see that Dante enters the higher spheres of Paradise through this 'door', his birth constellation. However, common wisdom surrounding the Barolo history takes the Southern Cross as Dante's 'gateway to the heavens'.[35] Why?

There is indeed a famous passage in *Purgatorio* related to 'four stars' which have been related to the constellation Crux:[36]

> The beauteous planet, that to love incites,
> Was making all the orient to laugh,
> Veiling the Fishes that were in her escort.
> To the right hand I turned, and fixed my mind

---

33 'Paradise', Canto XXII, 106–11.

34 Par. XXII, 112–17.

35 Hilger 'Capriccio italiano', p. 43.

36 Elly Dekker, 'The Light and the Dark: a reassessment of the discovery of the Coalsack Nebula, the Magellanic Clouds and the Southern Cross', *Annals of Science*, Vol. 47, no. 6 (1990), pp. 529–60.

> Upon the other pole, and saw four stars
> Ne'er seen before save by the primal people.
> Rejoicing in their flamelets seemed the heaven.
> O thou septentrional and widowed site,
> Because thou art deprived of seeing these!
> When from regarding them I had withdrawn,
> Turning a little to the other pole,
> There where the Wain had disappeared already.[37]

The 'beauteous planet, that to love incites' is Venus, which, in Dante's vision, dominated with its light the eastern sky and covered with its luminosity the constellation Pisces (the Fish) that escorted it. Dante had started his trip during Easter, namely, when the Sun lies in Aries (the Ram), according to the tropical zodiac. That Venus is visible in Pisces indicates that it is appearing as a morning star, rising ahead of the Sun, before dawn. The symbolism suggests that the pilgrims are enjoying the dawn of resurrection, Easter day, during the fourth day of their journey.[38]

**The Southern Cross?**
What does Dante mean by the statement that he 'saw four stars ne'er seen before save by the primal people'? It has been suggested that these are the stars of the Southern Cross, which Dante might have known, even if indirectly, from the *Almagest* where they appeared as part of the constellation Centaurus.[39] The possible source might have been Alfraganus' *Compendium of Astronomy*.[40] Dante certainly was aware of Alfraganus, and mentioned him at least once in *The Banquet*.[41] Moreover, can we infer from this that Palanti wanted 'his' Palace to incorporate some feature allegorically related to the Cross?

---

37 Pur. I, 19–30.

38 Dauphiné, *Le cosmos de Dante*, p. 48.

39 Claudius Ptolemy, *Almagest*, VIII.1. Constellation XLIV: Centaurus, trans. G. J. Toomer, (Princeton, NJ: Princeton University Press, 1998), pp. 394–96.

40 Bahrom Abdukhalimov, 'Ahmad Al-Farghani and his Compendium of Astronomy', *Journal of Islamic Studies*, Vol. 10, no. 2 (1999): 142–58.

41 *The Banquet*, II, Chapter XIV.

It might be that Dante 'invented' these stars right away, with no relation to an existing star catalogue or travelers' description whatsoever. To support this view, we have his explicit words 'four stars ne'er seen before save by the primal people' which might mean that only the inhabitants of the Garden of Eden, 'the primal people'—Adam and Eve—could have seen them, but *not* the rest of their descendants. Moreover, the poet just mentions the 'four stars' but says nothing whether these formed part of an astronomical asterism or constellation. It is not important whether Dante really knew of the Southern Cross's existence. What is important is that the notion that he did has long been believed.[42]

Palanti apparently wanted some celestial allegory to be attached to the highest—and more divine—part of the votive temple he built. The lighthouse, representing the nine angelical choruses surrounding God, with its powerful light that could be seen from the opposite shore of the River Plate, had certainly some role in guiding the European pilgrims towards the South, towards this promised land he might have thought Buenos Aires would be, and away from the Old Continent that would soon be devastated by a new world war. Eventually, as an astronomical guide for pilgrims and travelers heading south, the Cross was the most appropriate 'sign' in the sky.

However, if Palanti really wanted those powerful twin beacons, the Salvo Palace in Montevideo, and The Barolo, on the Argentine shore, to be related to Dante's entrance gate to the heavens, he probably should have thought more in Dante's 'glorious stars', especially in the *twins* Castor and Pollux placed in 'Leda's lovely nest', namely in Gemini, his birth sign, rather than in the Southern Cross; the latter being a constellation Dante probably never saw with his own eyes (Fig. 14).[43]

---

42 Charles Allen Dinsmore, *Aids to the Study of Dante* (New York: Houghton, Mifflin and Co, 1903), p. 240.

43 For Castor and Pollux see Pur. IV, 61.

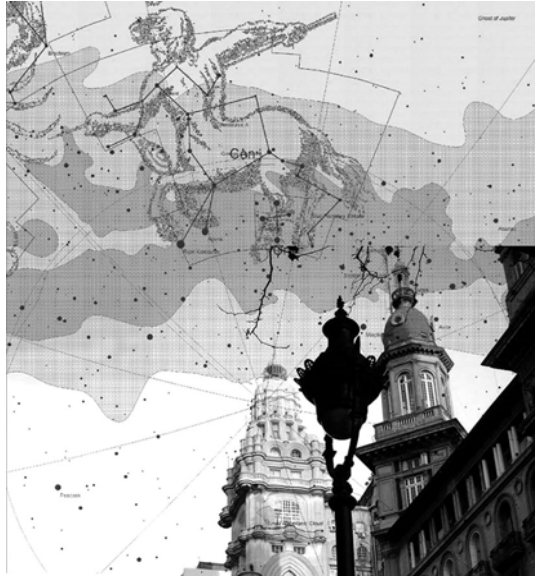

**Figure 14**: In spite of the fact that Gemini seems to be Dante's most probable constellation and birth sign (Par. XXII, 112–17), common wisdom surrounding *The Barolo*'s history takes the Southern Cross as Dante's 'gateway to the heavens'. Moreover, many sources (*e.g.*, Hilger, 'Capriccio italiano') suggest that the Cross is aligned with the ascensional axis of the building *just* on the first days of June, at precisely 19:30 Greenwich time.[44] Clearly, this is *not* true, as the 'alignment' takes place for a whole range of nights, provided the time of occurrence changes consistently day after day, of course. Furthermore, in the main text we suggest that the most appropriate constellation to be related with Dante's journey is not Crux, but rather Gemini. The image shows a photomontage (a crude composite photograph) of the Barolo Palace, together with the starry sky (including the South Celestial Pole and the Cross) for June 3rd at 23:45 Greenwich time (*i.e.*, 20:45 local time), for an observer in Buenos Aires.[45]

---

44 A *project* presented in March 2007—and approved in May 2007—in the Chamber of Deputies of Argentina (record 0164-D-2007, Commission for Culture), with an *expression of interest* for the restaurant on the ground floor of *The Barolo*, changes this to 19:45 without explicitly mentioning whether it is local or Greenwich times.

45 The starry sky shown in the picture is virtually the same also for June 18, at 22:45 Greenwich time. The sky 'rotates' 15 degrees around the South Celestial Pole in 1 hour. Moreover, for a fixed time of the day, the sky *shifts* roughly by 1

## Conclusion

Although Dante scholars have long discussed about the exact date of his journey through the three realms of the dead, nowadays there is an increasing consensus for the night before Good Friday in the year 1300.[46]

This is mainly because of its symbolic value and, no less relevant, because Dante himself hints to it in the *Inferno*. It coincides with the first Christian jubilee—with its 'great remissions and indulgences for sins'— proclaimed by Pope Boniface VIII, and this date is also consistent with the very first lines of the *Commedia*: 'Midway upon the journey of our life/ I found myself within a forest dark,/ For the straightforward pathway had been lost'.[47] Given that the Old Testament indicated a life expectancy of 70 (for example, 'The days of our years are seventy' in Psalm 90:10), and given that Dante was born in 1265, the 'midway upon the journey' of his life—that is, his 35 years of age—clearly coincided with the year 1300. Dante himself wrote that, 'I believe that in the perfectly natural man [the top of the arch of his life] is at the thirty-fifth year'.[48]

Palanti would, of course, not forget about this datum and, when selecting the actual location for the future construction of the building, the appropriate block of *Avenida de Mayo* was chosen. The Palace would be located at the 13th block of the *Avenida de Mayo* at number 1370. Construction began in 1919 but, in spite of the efforts and resources deployed, it was not finished in time. The building was inaugurated in 1923, a couple of years late for the sixth centenary of Dante's death. Luis Barolo, however, died in 1922, and could not assist to the official opening of July 7, 1923, with the blessing of the Apostolic Nuncio Msgr. Giovanni Beda Cardinale.[49]

---

degree each day (completing 360 degrees in roughly 365 days of the year). So, in 15 days (June 18 = June 3rd + 15 days) the night sky looks virtually the same as for June 3rd provided we look at it one hour earlier (*i.e.*, at 22:45 instead of at 23:45 Greenwich time).

46 Dauphiné, *Le cosmos de Dante*, p. 52 ; Battistessa 'Introducción', p. 38.

47 Inf. I, 1–3.

48 *The Banquet*, IV, Ch. XXIII.

49 Justo Solsona and Carlos Hunter, *La Avenida de Mayo. Un proyecto inconcluso* (Buenos Aires: Libreria Técnica CP67 S.A., 1990).

Paul Valéry once said that 'with the great poems happens the same as with Temples, in that they can still be admired even when the sanctuary is devoid of people and the feelings and causes which promoted their construction have faded away with time'.[50] Valéry, an agnostic, did not have a low opinion of beautiful cathedrals *just* because religion was the main cause for their construction. Dedicated efforts by many people ended up in a monumental recognition for the topography of Dante's verses and cosmology, which was shaped in reinforced concrete. *The Barolo* offers a vivid and prominent example of the role played by Italian culture and the medieval cosmos in modern times.

---

50 Cited in Battistessa, 'Introducción', p. 38.